\documentclass[twocolumn,showpacs,preprintnumbers,amsmath,amssymb]{revtex4}
\usepackage{epsfig}

\begin{document}

\title{The N(1710)~P$_{11}$ state is  confirmed in the re-analysis of the \mbox{$\pi$N $\rightarrow$ K$ \Lambda$} 
production; \\ it  is a good candidate for  a non-strange pentaquark}

\author{S. Ceci, A. \v{S}varc and B. Zauner}

\address{Rudjer Bo\v{s}kovi\'{c} Institute, \\
Bijeni\v{c}ka c. 54, \\ 
10 000 Zagreb, Croatia\\ 
E-mail: alfred.svarc@irb.hr}

\date{\today}

\begin{abstract}
Re-analyzing the old  \mbox{$\pi$N $\rightarrow$ K $\Lambda$} data, the additional proof is given for the existence 
of the N(1710)~P$_{11}$ state, critically needed  in light of reported observations of exotic  $\Theta$(1539) and 
$\Xi$(1862) pentaquarks. An existing  single-resonance model with S$_{11}$, P$_{11}$ and P$_{13}$ Breit-Wiegner 
resonances in the s-channel has been applied. It has been shown that the standard set of resonant parameters fails to 
reproduce the shape of the differential cross section. The new resonance parameter determination has been performed  
keeping in mind the most recent knowledge about  nucleon resonances.  The extracted set of parameters has confirmed 
the need for the strong contribution of a  N(1710)~P$_{11}$ resonance. The need for any significant contribution of 
the P$_{13}$ resonance has been eliminated. To reproduce the total cross section  at the same time with the linear 
dependence of the differential cross sections with the cos$\, \theta$ in the energy range 1650 MeV $<$ W $<$ 1800 MeV 
the P$_{11}$ resonance can not but be quite narrow.  
\end{abstract}  

\maketitle
The confirmation  and the additional proof for the existence of the N(1710)~P$_{11}$ resonance quite recently turned 
out to be critically needed  in light of reported observations of exotic pentaquark $\Theta$(1539) \cite{Pen104} and 
$\Xi$(1862) \cite{Pen204} states. 
The N(1710)~P$_{11}$ state is  associated with the non-strange member of the pentaquark anti-decuplet, and octet 
$\otimes$ anti-decuplet configurations predicted by chiral soliton \cite{Dia97,Dia03} and $qqqq \bar{q}$ strong 
color-spin correlated models \cite{Jaf03}.  For fixing the scale, the chiral soliton model \cite{Dia97,Dia03} has 
used  the mass of the  N(1710)~P$_{11}$ resonance as the non-strange pentaquark input. The $qqqq \bar{q}$ strong 
color-spin correlated model \cite{Jaf03}  has fixed its scale by identifying the the first - (${\bf [ud]^2 \bar{s}}$) 
state as $\Theta^+$, the second - (${\bf [ud]^2 \bar{d}}$) state as N(1440)-Roper resonance, and the  
N(1710)~P$_{11}$ state is identified with the third - (${\bf [ud] [su]_+ \bar{s}}$) hidden strangeness state N${_s}$. 
As the exotic pentaquark states turn out to be experimentally established, it is reasonable to assume that the 
physical  N(1710)~P$_{11}$ state corresponds to a pentaquark state as well. The main aim of this paper is, therefore, 
the confirmation of the N(1710)~P$_{11}$ existence. The known problem of the discrepancy in width remains 
unaddressed: when accepted,  the N(1710)~P$_{11}$ physical state is fairly wide ($90<\bar{\Gamma}_{\rm PDG}< 
480$~MeV) \cite{PDG}, but both quark models have predicted that the  width of a non-strange member of the pentaquark 
multiplet is narrow ($\Gamma _{\rm pq} <$~40~MeV). 

At this moment it is important to emphasize that without the $K \Lambda$ data, even the sole existence of  
N(1710)~P$_{11}$ resonance  is  seriously questioned \cite{Arn85,Arn95,Arn04}. The energy-dependent coupled-channel 
Chew-Mandelstam K matrix methods do not see any N(1710)~P$_{11}$ state, while all coupled-channel T-matrix 
Carnegie-Mellon Berkeley type models (CMB) \cite{Cut79,Bat98,Vra00}, however, do indisputably  need it, and report it 
to be  strongly inelastic. In the only coupled-channel PWA which includes the $K \Lambda$ data \cite{Feu98} the poles 
and residues are extracted on the basis of speed plot technique described by H\"{o}hler \cite{Hoh93}, and the direct 
calculation of the T-matrix in the complex energy plane is bypassed because of technical complexity needed for the 
analytical continuation  of all Feynman diagrams into the complex energy plane.  Anyhow, it comes out with the result 
that the N(1710)~P$_{11}$ resonance definitely exists, and exhibits a strong coupling to inelastic channels, $K 
\Lambda$ in particular. Let us point out that the problem of identifying the N(1710)~P$_{11}$ state seems to be more 
of fundamental than of technical nature. Namely, in the paper by Cutkosky and Wang \cite{Cut90} it has been shown 
that the coupled-channel method predicts the existence of additional N(1710)~P$_{11}$  resonance in the energy range 
where the energy-dependent coupled-channel Chew-Mandelstam K matrix method does not see anything, when  fitting the 
identical set of single energy T-matrices. It has been explicitly concluded that the differences in resonance 
structure in the two afore described models arise from the different parameterization of the energy dependence, 
rather than differences in the data. The answer to this puzzle is yet to be given, but it is conceivable  that the 
number of imputed channels is insufficient, and that more rigorous inclusion of $K \Lambda$ channel is needed.

The experimental data for the process \mbox{$\pi$N $\rightarrow$ K$ \Lambda$} are available for quite some time 
\cite{Dyc69,Jon71,Kna75,Bak78}. The  different measurements have agreed just upon a few issues: the data  clearly 
show a distinct structural behavior in the vicinity of 1700 MeV and  the differential cross sections 
\cite{Kna75,Bak78} show a clearly recognizable linearity in cos$\, \theta$ up to 1850 MeV. As the experiments admit 
the systematic error of 8 - 15 \% in absolute normalization,  the observed structure is smeared in  energy making any 
conclusions about the profile of the structure very difficult. Even when  the agreement about the position and the 
width of the structure is achieved, the  interpretation of its origin remains unclear.  It can be interpreted either 
as a genuine T-matrix pole, namely the signal of a $qqq$ or a $qqqq \bar{q}$ resonance, or as a cusp effect resulting 
from the opening of the $K \Sigma$ channel. The distinction between the resonant and the cusp effect interpretations 
is as well  non-trivial. To claim that the structure is a genuine T-matrix pole requires a full scale coupled-channel 
analysis which manifestly incorporates the appearance of cusp effects, and a clear and unambiguous search for the 
poles in the complex energy plane via well defined analytical continuation. Anything less (Argand diagrams, fits to 
different single-resonance models,..) fails on the basis of first principles as it has been demonstrated in the 
search for the dibaryon resonance hidden in the $^1$D$_2$ and $^3$F$_3$ partial waves in the pp $\rightarrow$ pp 
elastic scattering \cite{Loc86}. 
Partially because of data dissipation, and partly because of technical complexity, the \mbox{$\pi$N $\rightarrow$ K$ 
\Lambda$} process  have not yet been included as the part of the data base in the existing coupled-channel 
$\pi$-nucleon partial wave analyses (PWA) \cite{Arn85,Arn95,Arn04,Cut79,Bat98,Vra00,Man84,Man92,KH84,Gre99} until 
late 90-es \cite{Feu98}, and even there, according to the authors themselves,  play a minor role because of large 
errors and are included for completeness only. 

The N(1710)~P$_{11}$ resonance parameters have been for the first time extracted from the  \mbox{$\pi$N $\rightarrow$ 
K$ \Lambda$} data in a single-channel energy-dependent phase shift analysis \cite{Bak77}, which was soon afterwards 
upgraded using the new set of data \cite{Bak78}. Because of using only one channel, the extracted resonances for the 
S$_{11}$, P$_{11}$ and P$_{13}$  partial waves do not necessarily  describe all $\pi$-nucleon channels at the same 
time, so the single-channel result should be coordinated with the values obtained in the multi-channel PWA. That has 
been attempted in ref. \cite{Feu98}. 

  In this article the differential cross section for the \mbox{$\pi$N $\rightarrow$ K$ \Lambda$} process is described 
as a coherent sum of Breit-Wigner resonance contributions for the s-channel:
\begin{eqnarray} 
\frac{{\rm d} \sigma}{{\rm d}\Omega}&=& N \left| \sum_{l=0}^{N}a_{l^\pm}T_{2I,2l^{\pm}}^{\pi N,K \Lambda}\right| ^2 \ 
; \ \ \ l^\pm=l\pm \frac{1}{2} \ , \nonumber \\
T_{2I,2J}^{\pi N,K \Lambda}&=& \sqrt{T_{2I,2J}^{\pi N,\pi N} \cdot \ \ 
T_{2I,2J}^{K\Lambda,K\Lambda} }, \nonumber \\
T_{2I,2J}^{ch,ch}&=&\frac{x_{ch} \frac{\Gamma}{2}}{M-{\rm W} -i\frac{\Gamma}{2}} \ .
\label{eq:sres}
\end{eqnarray}
where N is the normalization constant, $a_{l^\pm}$ are corresponding angular dependent expansion coefficients while 
$M$, $\Gamma$ and $x_{ch}$  are resonance masses, widths and branching ratios respectively. The  t-channel resonant 
contribution of $K^*$(892) meson is in this calculation neglected, because the product of the $K^* \pi K $ and $K^* 
\Lambda N$ coupling constants, which enters the model, is poorly determined, and due to the kinematical behaviour of 
the $K^*$(892) propagator influences the shape of the differential cross section significantly only at energies far 
above the domain of interest of 1650-1750 MeV \cite{Feu98}. The square root recipe for the $T_{2I,2J}^{\pi N,K 
\Lambda}$ matrix is generally valid in a single resonance approximation, but is believed to be a fair approximation 
within a resonance energy domain for any full calculation. Therefore, the use of a single resonance model should be 
sufficient to establish the relevant resonance parameters near  the top of the resonance. 

The collection of data sets which we have used for a comparison with the predictions of our model 
\cite{Dyc69,Jon71,Kna75,Bak78} declare an 8 - 15 \% systematic error in the absolute normalization, therefore an 
overall normalization of absolute scale is in order. The smearing of the structure in the total cross section, due to 
the normalization uncertainties in the reported experiments has been lessened by creating the amalgamated data set. 
Because three out of four experiments \cite{Dyc69,Kna75,Bak78} report the peak of  920 $\mu$b at the energy of 
1694~MeV, we have chosen to normalize all data accordingly.  In addition, we have shifted the whole Knasel et al.  
data set  \cite{Kna75}  up in energy for 9 MeV. That step is open for criticism because it implicitly questions the 
energy calibration of that particular experiment. That is not our intention. At this point, we are primarily 
interested in the shape of the structure, namely its width, because we tend to interpret it as a possible narrow 
non-strange pentaquark candidate. The mass of the resonance is not of our prime conceren. By shifting the peak of the 
structure to the same energy we automatically extract the common width from all experiments, because otherwise the 
width is smeared and comes out bigger then actually extracted from each experiment separately. We could have shifted 
other three experiments 9 MeV down in energy and obtain  the same width, but lower mass. 

The single-resonance model calculation has been repeated  using the standard set of parameters for the  
N(1650)~S$_{11}$, N(1710)~P$_{11}$ and N(1720)~P$_{13}$ resonances of ref. \cite{PDG},  the parameters  are given in 
Table 1 denoted as "PDG", and the agreement with experiment is given in Figs. 1. and 2. as thin solid line. 
That choice of parameters reproduces  the absolute value of the total cross section reasonably well, and manages to 
reproduce the shape of the angular distribution  only at W=1683 MeV. However, it fails miserably in reproducing the 
shape of the differential cross section data at other energies. 

\begin{table}[!h]
\caption{ Resonance parameters for the single resonance  model.}
{\scriptsize
\begin{tabular}{ccccc}
\hline \hline
{} & {$M[MeV]$}&$\Gamma[MeV]$ &$x_{\pi N}$[\%]& $x_{K \Lambda}$[\%]\\ 
  \cline{2-5}
{} & $S_{11}$ {} $P_{11}$ {} $P_{13}$&$S_{11}$ {} $P_{11}$ {} $P_{13}$&$S_{11}$ {} $P_{11}$ {} $P_{13}$&$S_{11}$ {} 
$P_{11}$ {} $P_{13}$\\ 
\hline \hline 
{} &{} &{} &{} &{} \\ [-1.5ex]
$PDG$ &1650 {}   1710 {}  1720 &150   {} 100  {}   150 &70   {   }  15   {   }  15    &7.0    {}   15   {}   6.50  \\ 
$Sol  \ 1$ &1652  {} 1713  {}  1720 &202  {}   180 {}  244& 79 {  }  22  {  }  18 &{\bf 2.4}   {}  {\bf 23} {}  {\bf 
0.16}\\ 
$Sol  \ 2$ &1652 {}  1713  {}  1720 &202  {}   180  {}  244& 79 {  }  22  {  }  18 & {\bf 2.4}  {}   {\bf 35}  {}  
{\bf 0.16}\\ 
$Sol  \ 3$ &1652 {}  {\bf 1700} {}   1720 &202  {}   {\bf 68}  {}  244& 79 {  }  22 {  }   18  & {\bf 3.0}  {}   {\bf 
32}  {} {\bf 0.16} \\ 
\hline \hline
\end{tabular}
    }
\label{table1} 
\end{table}
The disagreement of the calculated shape of the differential cross section with the amalgamated data set has been 
eliminated by performing a fit to the amalgamated data set in such a way that we have kept masses and widths of the 
N(1650)~S$_{11}$, N(1710)~P$_{11}$ and N(1720)~P$_{13}$ resonances fixed to the values given in ref. \cite{Bat98}, 
and varied the  $K \Lambda$ branching ratios in the Breit-Wigner parameterization given in Eq.(\ref{eq:sres}). The 
energy dependence of the Breit-Wigner resonance width, which ensures the correct threshold and high energy behavior 
is kept as in ref. \cite{Bat98}. 

We have obtained a good description of the  absolute value of the total cross section  maintaining  the  shape of the  
differential cross sections of the \mbox{ $\pi$N $\rightarrow$ K$ \Lambda$} process, which turnes out to be  linear 
in cos$\, \theta$ (indicated by experimental data of ref. \cite{Kna75,Bak78}). This has technically been done by 
dividing the total $\chi ^2$ into two equally contributing sub parts: one originating from the total, and second from 
the differential cross sections. 
\begin{figure}[!h]
\label{fig:1}
\epsfig{figure=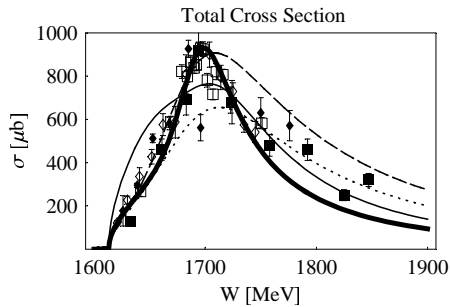,height=4cm,angle=0}
\caption{The agreement of the amalgamated experimental data for the total cross section (ref. 
\protect\cite{Dyc69}-full diamonds; ref. \protect\cite{Jon71}-open diamonds; ref. \protect\cite{Kna75}-open boxes and 
ref. \protect\cite{Bak78} full boxes) with the single resonance model predictions using different inputs for the 
resonance parameters: "standard" (PDG) set (thin solid line); $Sol \ 1$ (dotted line); $Sol \ 2$ (dashed line) and 
$Sol \ 3$ (thick solid line).}
\end{figure}
The special care has been taken to reproduce the peak experimental data at W~$\approx$ 1700~MeV because the validity 
of the single resonance model is expected to weaken outside that range. In that way we have obtained $Sol \ 1$ and 
$Sol \ 2$ with the Breit-Wigner parameters given in Table 1, and the agreement with data  given with dotted and 
dashed lines in Figs. 1. and 2. 

Because of the observed linearity of the differential cross section, the contribution of the N(1720)~P$_{13}$ 
resonance notably decreases. 

Consequently, due to the reduction of the total cross section, the contribution of  the $P_{11}$ resonance to $K 
\Lambda$ channel  significantly rises.

The branching ratio of $S_{11}$ resonance to $K \Lambda$ channel turns out to be somewhat  smaller then previously 
believed.

Calculated total \mbox{ $\pi$N $\rightarrow$ K$ \Lambda$}   cross sections ($Sol \ 1$ and $Sol \ 2$) {\it can not} 
describe amalgamated  data set from  1650 to 1800~MeV successfully for any choice of $K \Lambda$ branching fraction 
for the resonance masses and widths of ref. \cite{Bat98}.  The $Sol \ 1$ fits the upper part of the energy range and 
under-shoots the peak of the resonance, while $Sol \ 2$ fits the resonance part but over-shoots the high energy part. 

That conclusion is in complete agreement with the latest description of the \mbox{ $\pi$N $\rightarrow$ K$ \Lambda$} 
total cross section data given in ref. \cite{Pen02}. If the constructed amalgamated data set gets the experimental 
confirmation the N(1710)~P$_{11}$  resonance can not be as wide as presently expected \cite{PDG}.

In order to improve the agreement of the single resonance model with the amalgamated data set, namely to generate a 
solution which agrees with the total cross section throughout the whole energy range simultaneously keeping the 
linearity of differential cross section we have performed a fit where we have released the width of the 
N(1710)~P$_{11}$  resonance. We have obtained $Sol \ 3$, given in Table 1, and compared with experimental data in 
Figs. 1. and 2. (thick solid line). We have obtained the best agreement with the experiment  for a {\em strongly 
inelastic  and quite narrow} $P_{11}$ resonance. 
\begin{figure}[h]
\label{fig:2}
\epsfig{figure=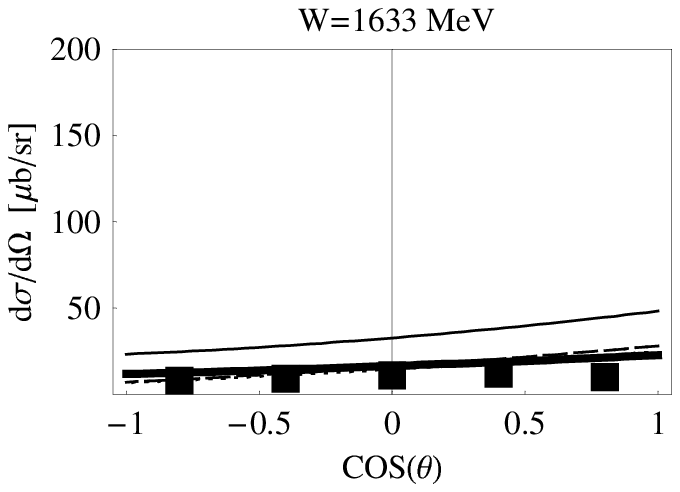,height=2.4cm,width=2.8cm,angle=0}
\epsfig{figure=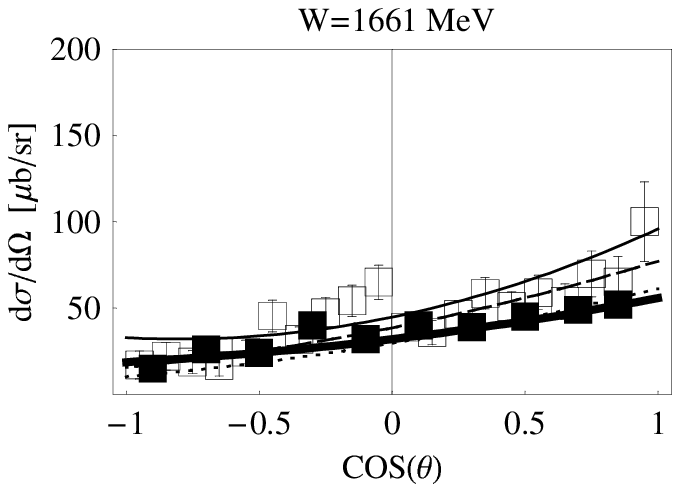,height=2.4cm,width=2.8cm,angle=0}
\epsfig{figure=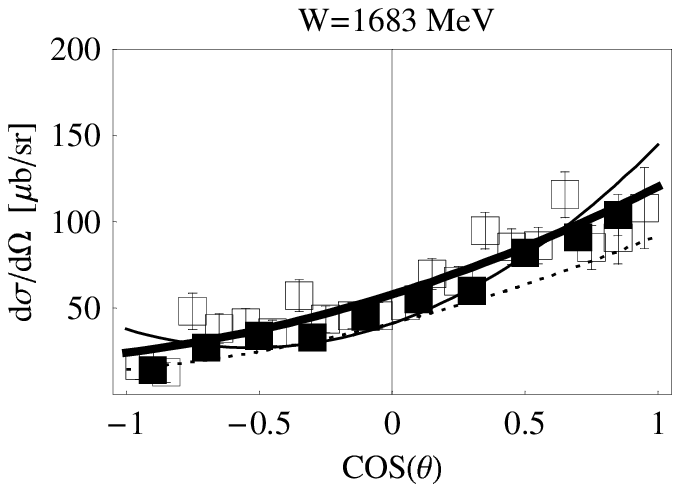,height=2.4cm,width=2.8cm,angle=0} \\
\epsfig{figure=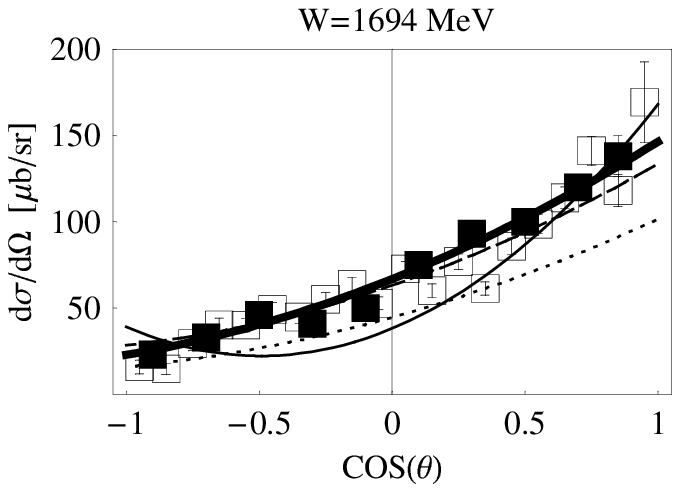,height=2.4cm,width=2.8cm,angle=0}
\epsfig{figure=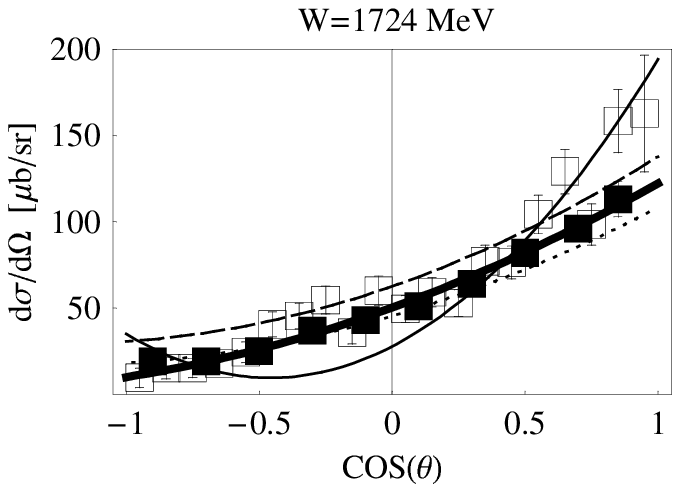,height=2.4cm,width=2.8cm,angle=0}
\epsfig{figure=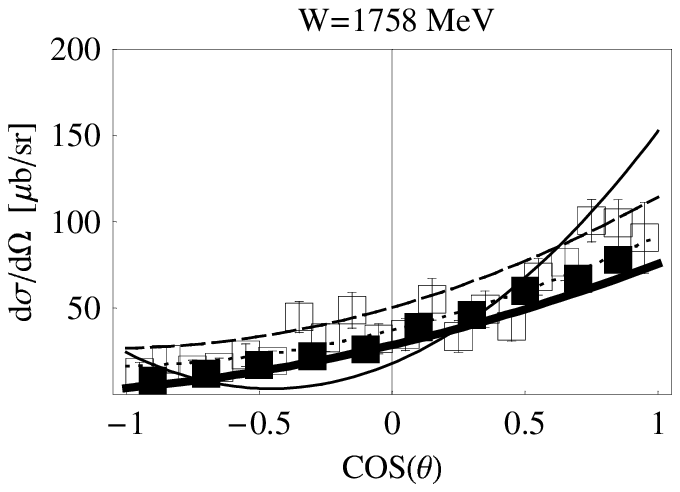,height=2.4cm,width=2.8cm,angle=0} \\
\epsfig{figure=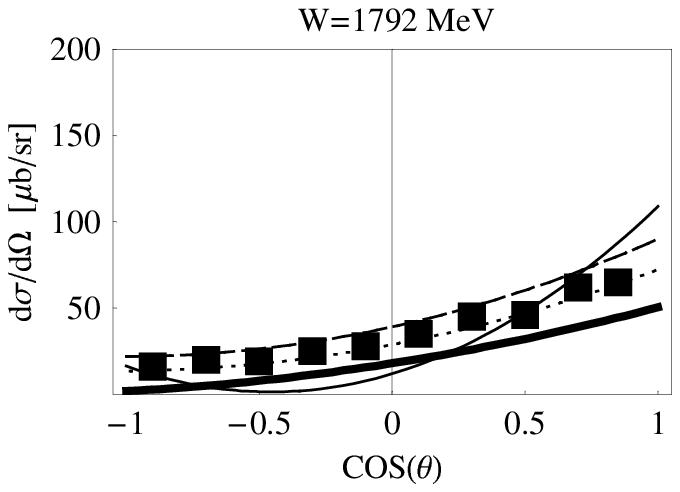,height=2.4cm,width=2.8cm,angle=0}
\epsfig{figure=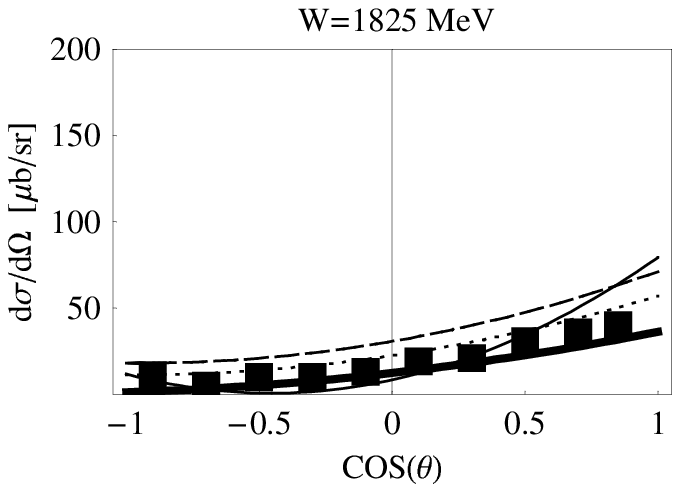,height=2.4cm,width=2.8cm,angle=0}
\epsfig{figure=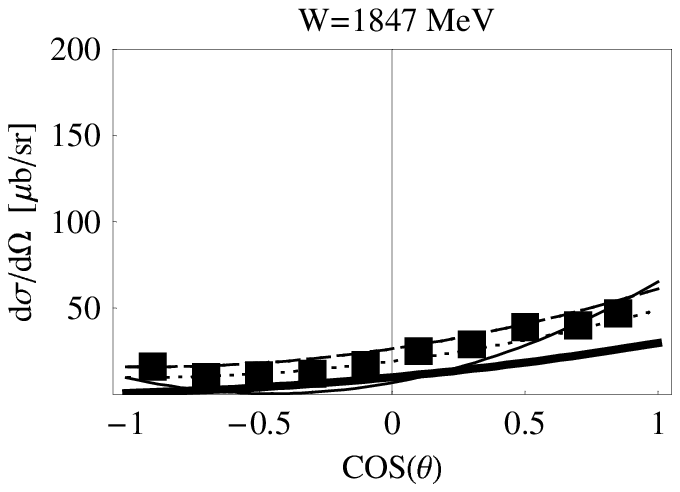,height=2.4cm,width=2.8cm,angle=0} 
\caption{The agreement of the amalgamated experimental data for the differential cross section (ref. 
\protect\cite{Kna75} - open boxes and ref. \protect\cite{Bak78} - full boxes); with the single resonance model 
predictions using different inputs for the resonance parameters: "standard" (PDG) set (thin solid line); $Sol \ 1$ 
(dotted line); $Sol \ 2$ (dashed line) and $Sol \ 3$ (thick solid line).}
\end{figure}
The conclusion about the width of the N(1710)~P$_{11}$  resonance is valid only under the condition that its 
branching fraction  to the $K \Sigma$ channel is small, otherwise the cusp effect might effectively reduce the width 
obtained within the framework of the single resonance model. However, according to the ref. \cite{PDG}, that 
condition seems to be accomplished.
 
Believing that the necessary conditions are fulfilled we conclude that the narrow P$_{11}$  state ($\Gamma \approx 
68$~MeV) with the mass of 1700 MeV is not only consistent, but even favored by the existing data set, in particular 
when the latest measurement giving not only total cross section but angular distributions as well \cite{Bak78} is 
taken as most reliable.
 
The tendency of having  the width of the  N(1710)~P$_{11}$ state narrower in $K \Lambda$ single-channel analysis  
then in other $\pi$-nucleon channels \cite{PDG} offers the speculation that in that particular channel we do see the 
admixture of the pentaquark state to a standard  $qqq$ state. We illustrate that statement.

A simple non-unitary addition of extra P$_{11}$ resonance to a single resonance model (now we have {\it two} 
resonances with the strong $K \Lambda$ branching ratio in the P$_{11}$ partial wave) further improves the agreement 
of the model with experiment. The fit gives a solution with both P$_{11}$ resonances degenerated in mass (M $\approx$ 
1700 MeV), but the first P$_{11}$ resonance tends to be much narrower ($\Gamma \approx$~40~MeV), while the second one 
is much wider ($\Gamma \approx$~240~MeV).  However, having an "extra" P$_{11}$ state is not altogether a novelty, as 
a second P$_{11}$ state at 1740$\pm$11 MeV is reported to be seen in the PWA analysis of ref. \cite{Bat98} in 1995. 
overall 

In order to confirm that the N(1710)~P$_{11}$   resonance is, or at least has the admixture of a non-strange 
pentaquark,  the precise measurement of the total cross section and angular distributions of the $\pi$N $\rightarrow$ 
K$ \Lambda$ process in the energy range \mbox{1613 MeV $<$ W $<$ 1900 MeV} is urgently needed. The non-ambiguous 
elimination of a cusp effect as a source of the observed structure  can be given only in the framework of a 
coupled-channel, multi-resonance theoretical analysis. As the  the $K \Sigma$ channel is the most probable reason for 
the cusp effect it has to be included in such an analysis. The data for the $\pi$N $\rightarrow$ K$ \Sigma$ process 
in the energy range \mbox{1683 MeV $<$ W $<$ 1900 MeV} are scarce \cite{KSigma} so the measurement of the total cross 
section and angular distributions  for that process are badly needed as well.
In this article, using  the single resonance model only, we  show a strong indication that a standard, wide P$_{11}$ 
resonance  is incompatible with the existing data, and that a narrow P$_{11}$ state weakly coupled to the $K \Sigma$ 
channel is quite probable. 
 \\ \\ \\
\noindent
{\bf Summary} \\
The single resonance model calculation using the standard set of parameters for the  N(1650)~S$_{11}$, 
N(1710)~P$_{11}$ and N(1720)~P$_{13}$ resonances of ref. \cite{PDG} reproduces  the absolute value of the total cross 
section reasonably well, and manages to reproduce the shape of the angular distribution  only at W=1683 MeV. However, 
it fails miserably in reproducing the shape of the differential cross section at other energies. 

The single resonance model fits indicate that the linear shape of the differential cross section can be maintained 
only if the contribution of the N(1720)~P$_{13}$  resonance is negligible, and that the agreement of the total cross 
section with the data in the energy range \mbox{1650 $<$ W $<$ 1800 MeV} can be achieved only if the N(1710)~P$_{11}$ 
resonance is narrow ($\Gamma \approx$ 68~MeV).

In order to give any conclusions about the nature of the 1700 Mev structure, the re-measuring of the total and 
differential cross sections for the \mbox{$\pi$N $\rightarrow$ K$ \Lambda$} and \mbox{$\pi$N $\rightarrow$ K$ 
\Sigma$} processes  in the energy range \mbox{1613 MeV $<$ W $<$ 1800 MeV} is badly needed because of the serious 
incoherence of experimental data.
The decisive conclusion about the existence of the $P_{11}$ non-strange pentaquark will be possible only when the 
improved set of data is fully incorporated in one of the existing coupled-channel partial wave analyses 
\cite{Arn85,Arn95,Arn04,Cut79,Bat98,Vra00,Man84,Man92,KH84,Gre99} .
\\ 
\noindent
{\bf Acknowledgment} \\
We are grateful to K. Kadija from  Rudjer Bo\v{s}kovi\'{c} Institute, a member of Na49 Collaboration, for a constant 
encouragement to link our "P$_{11}$" problems with the new pentaquark states. We also thank M. Korolija from the same 
Institute for initiating out interest in $\pi$N $\rightarrow$ K$ \Lambda$ process.

\end{document}